\documentclass[review]{elsarticle}
\usepackage{amsmath,bm}
\usepackage{graphicx}
\usepackage{amsmath}
\usepackage{amsthm}
\usepackage{float}
\usepackage{epsfig,graphics,subfigure,psfrag,amsmath,amssymb}
\usepackage{pdfpages}
\usepackage{subfigure}
\usepackage{graphicx}
\usepackage{booktabs}
\usepackage{moreverb}

\usepackage{multirow} 
\usepackage{amsmath}
\usepackage{xcolor}
\usepackage{algorithmic}
\usepackage[vlined,boxed,ruled]{algorithm2e}
\newcounter{thm}
 \newcounter{ex}
 \newcounter{re}
\usepackage{lineno,hyperref}
\modulolinenumbers[5]

\journal{Journal of \LaTeX\ Templates}

\begin{document}

\begin{frontmatter}

\title{LPTD: Achieving Lightweight and Privacy-Preserving Truth Discovery in CIoT}




\author[a]{Chuan Zhang}

\author[a]{Liehuang Zhu}

\author[a]{Chang Xu\corref{mycorrespondingauthor}}

\cortext[mycorrespondingauthor]{Corresponding author}
\ead{xuchang@bit.edu.cn}

\author[a]{Kashif Sharif}

\author[b]{Xiaojiang Du}

\author[c]{Mohsen Guizani}

\address[a]{Beijing Engineering Research Center of Massive Language Information Processing and Cloud Computing Application,School of Computer Science and Technology, Beijing Institute of Technology, Beijing, China.}

\address[b]{Department of Computer and Information Sciences, Temple University, Philadelphia, USA.}

\address[c]{Department of Electrical and Computer Engineering, University of Idaho, Moscow, Idaho, USA.}

\begin{abstract}
	In recent years, cognitive Internet of Things (CIoT) has received considerable attention because it can extract valuable information from various Internet of Things (IoT) devices. 
	In CIoT, truth discovery plays an important role in identifying truthful values from large scale data to help CIoT provide deeper insights and value from collected information. However, the privacy concerns of IoT devices pose a major challenge in designing truth discovery approaches. Although existing schemes of truth discovery can be executed with strong privacy guarantees, they are not efficient or cannot be applied in real-life CIoT applications. This article proposes a novel framework for lightweight and privacy-preserving truth discovery called LPTD-I, which is implemented by incorporating fog and cloud platforms, and adopting the homomorphic Paillier encryption and one-way hash chain techniques. This scheme not only protects devices' privacy, but also achieves high efficiency. Moreover, we introduce a fault tolerant (LPTD-II) framework which can effectively overcome malfunctioning CIoT devices.	
	Detailed security analysis indicates the proposed schemes are secure under a comprehensively designed threat model. Experimental simulations are also carried out to demonstrate the efficiency of the proposed schemes.
\end{abstract}


\end{frontmatter}

\section{Introduction}
Cognitive Internet of Things (CIoT) is a specialized IoT model which capitalizes on the  increasing capabilities of mobile devices (with built-in comprehensive sensor sets), which uses cognitive computing techniques to find valuable information from large scale sensing data \cite{WuDXFDWL14,MishraLC15,FengSH17}. By analyzing the big data created by various IoT devices, CIoT is able to provide deeper insights, high-level intelligence, and further create values for people. 

Despite the proliferation of CIoT, there are some increasing concerns which may impede its wide adoption. For example, the sensory data captured and provided by different devices is usually not directly usable or reliable, as it may be distorted due to reasons such as, lack of sensor calibration, poor sensor quality, background noise, and even the intent to deceive. Therefore, an important task of the CIoT applications is to discover truthful information from the sensory data. This task, called  truth discovery, has drawn significant attention \cite{LiLGZFH14,LiDLMS12,LiLGSZFH15}. Typically, the common principle to execute truth discovery is weighted aggregation that assigns a higher weight to a particular device if data reported by it is closer  to the aggregated results from all devices. Moreover, a device's data is given higher value if the device has higher weight due to its past performance \cite{MiaoJSLGQXGR15, XuLTLDY17}. By performing truth discovery, accurate sensory data can be obtained, and such data will greatly promote the effectiveness of CIoT applications.

Although having significantly improved data accuracy, the challenge for truth discovery, is that the sensory data is highly sensitive and should be well protected, especially considering that sensory data may contain personal information \cite{XiaoRSDHG07,DuC08,DuXGC07}. For example, geo-tagging services can publish timely and accurate localization of specific objects (e.g., pothole, automated external defibrillator, litter, etc.). However, this may lead to exposure of participating users' sensitive geo-location and/or movement patterns. Aggregated health statistics (i.e., treatment outcomes)  may provide valuable information regarding medical devices' effects or new drugs, but may threaten the privacy of participating patients. Meanwhile, user reliability (i.e., weight) is another private information which should be well protected. From user reliability information, the attacker may infer details of participating users' education, skills, and personality traits. For example, aggregating opinions regarding challenging social problems may lead to a better solution. However, the leakage of reliability may disclose users' education and intellectual level. 

Several studies have tried to preserve users' privacy in the applications of truth discovery \cite{MiaoJSLGQXGR15, XuLTLDY17, MiaoSJLT17}. However, most of them are not efficient or cannot be applied in real-life CIoT applications. For example, Du et al\cite{du2009transactions} tried to find a reliable key management scheme, Miao et al. \cite{MiaoJSLGQXGR15} proposed a cloud-based privacy-preserving truth discovery scheme to protect users' sensory data. However, by using threshold Paillier cryptosystem \cite{CramerDN01}, their scheme is not efficient. To improve efficiency, Xu et al. \cite{XuLTLDY17} proposed a lightweight and privacy-preserving discovery scheme by using the additive homomorphic privacy-preserving techniques.  Miao et al. \cite{MiaoSJLT17} further designed a lightweight truth discovery framework by using two non-colluding cloud platforms. Although their schemes achieve better efficiency, they cannot be applied in CIoT applications, especially in scenarios where some IoT devices may not deliver their data timely \cite{LuHLG17}. Moreover, all the above schemes
cannot defend from external attackers who inject false data into the system. Hence, there is a need for an efficient truth discovery scheme, which not only protects users' privacy, but is also able to mitigate false data injection attacks and give fault tolerance.

In this paper, to address these challenges, we present a lightweight privacy-preserving truth discovery scheme in CIoT, called LPTD-I, to protect devices' privacy (i.e., sensory data and reliability information), and resist false data injection attacks. The framework is implemented by involving fog and cloud platforms, adopting homomorphic Paillier encryption, and one-way hash chain techniques. In this framework, the fog node authenticates the data submitted from devices and aggregates the data before delivering it to the cloud. In addition, we exploit the properties of modular arithmetic to design a data aggregation algorithm which is efficient and privacy preserving.

Although LPTD-I can defend against the false data injection attack launched by external attackers, it is not fault-tolerant. Thus, we exploit the modified Paillier cryptosystem and propose a framework (LPTD-II) suitable for the scenarios where some IoT devices may stop delivering data due to device failure, to the fog node. In this framework, the secret key is split into two parts, and the fog devices can cooperate with the cloud to recover the aggregated results successfully. 

In summary, the contributions of this paper are:
\begin{itemize}
	\item We propose a novel lightweight and privacy-preserving truth discovery scheme in CIOT, called LPTD-I. This scheme not only preserves the privacy of users (i.e., sensory data and reliability information), but also achieves high efficiency.

	\item For the scenarios where some IoT devices stop reporting sensory data to the fog node, an upgraded technique called LPTD-II,  is proposed to achieve fault tolerance.
	
	\item Detailed security analysis indicates the proposed schemes are secure under an elaborate threat model. Additionally, experimentation shows the efficiency of both the proposed schemes.
\end{itemize}

The rest of this paper is organized as follows. In section 2, we give the problem definition which includes the system model, security model, and design goals. In section 3, we describe some preliminary. The details of the proposed LPTD schemes are described in section 4, followed by the security analysis and performance analysis in section 5 and section 6, respectively. In section 7, we discuss the related work. Finally, we draw the conclusion in the last section.

\section{Problem Definition}
The system model, security model, and design goals are outlined in the following sections.

\subsection{System Model}
\begin{figure}[htb]
	\centering
	\includegraphics[width=0.8\textwidth]{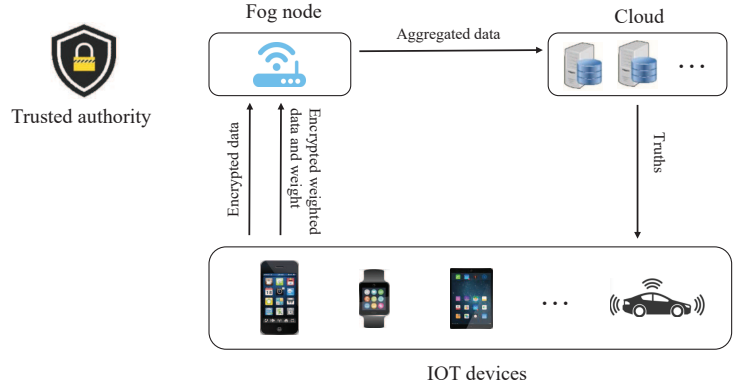}
	\caption{System model.}
	\label{fig:fs}
\end{figure}

The system model shown in Fig.~\ref{fig:fs} is comprised of four entities: IoT devices, the fog node, the cloud, and a trusted authority.
\begin{itemize}
	\item {IoT devices:} Each IoT device is equipped with sensing, communication, and computing capabilities, which can enable the device to collect sensory data, report data, and perform simple computation operations. Note that, since most IoT devices are resource-constrained, the computational costs for operations performed at these devices should be minimal.
	
	\item {Fog node:} The fog node acts as a middle layer between the IoT devices and the cloud, and is deployed at the edge of network. They can process/deliver data for the devices and/or cloud. In our schemes, it also aggregates all reports from IoT devices, and forwards resulting data to the cloud.
	
	\item {Cloud:} It receives all data from the IoT devices through the fog node. For each object, it generates an initial ground truth, and iteratively updates the truth in cooperation with the fog node.
	
	\item {Trusted authority (TA):} TA is a trusted third party, and it bootstraps the whole system. It generates keys and assigns them to all entities. Once the system is up and running, the TA remains offline.
\end{itemize}

We formalize the truth discovery approach as follows: Suppose there are $K$ IoT devices and $M$ objects, we use $x^k_m$ to denote the observed value of device $k$ for object $m$. For all devices, $\{w_1, w_2, \cdots, w_K\}$ are used to denote their reliabilities (i.e., weights). Each object is assigned an initial ground truth. The goal of the proposed scheme is to calculate the ground truths $\{{x^*_m}\}^{M}_{m=1}$ for all objects while protecting the observed value and weight of each device from being disclosed to others. Table \ref{tq} summarizes the main notations used in this work.

\begin{table}[htb]
	\caption{Summary of notations}.
	\begin{center}
		\small
		\begin{tabular}{|c|c|}
			\hline
			Symbol & Definition \\ \hline
			$K$ & Number of devices \\ \hline
			$k$ & Index of devices, $k \in \{1,K\}$\\ \hline
			$w_k$ & Weight of device $k$ \\ \hline
			$M$ & Number of objects \\ \hline
			$m$ & Index of objects, $m \in \{1,M\}$ \\ \hline
			$x^k_m$ & Observed value of device $k$ for object $m$ \\ \hline
			$x^*_m$ & Truth for the object $m$ \\ \hline
			$std_m$ & The standard deviation for the $m$-th object \\ \hline
		\end{tabular}
	\end{center}
	\label{tq}
\end{table}

\subsection{Security Model}
\begin{itemize}
	\item TA is considered to be fully trusted, and it cannot be breached by any attacker.
	
	\item The fog and cloud elements are honest-but-curious. This means that they will follow the protocol, but are also curious regarding device/user details. Note that, in our threat model, they do not collude with each other.
	
	\item The honest-but-curious IoT devices will follow the protocols. They can collude with other entities (i.e., other IoT devices, the fog, and the cloud), but we emphasize that they cannot collude with the fog and the cloud simultaneously.
	
	\item Since the focus of this work is to design a privacy-preserving truth discovery approach, internal attacks are not considered, i.e., all entities cannot be compromised at the same time. However, we do allow that some IoT devices may malfunction or stop reporting data intermittently. Moreover, external attackers may also launch false data injection attacks. Hence, the fog node should filter such data before transmitting them to the cloud.  
\end{itemize}

\subsection{Design Goals}
The goal of the proposed scheme is to design an efficient and privacy-preserving truth discovery approach which can protect devices' privacy and reduce computational costs. Security issues as studied in \cite{wu2014mobifish,wu2014security,huang2014achieving} should be solved in our work. In order to achieve this, following design goals must be guaranteed:

\begin{itemize}
	\item Privacy: The proposed scheme should preserve the privacy. The fog node and cloud can obtain the truthful values, but they cannot obtain individual IoT devices' information (i.e., sensory data and reliability information).
	
	\item {Security:} The scheme should be resistant to false data injection attacks launched by external attackers. In other words, the fog node should authenticate the IoT devices and filter the false data before transmitting it to the cloud.
	
	\item{Fault Tolerance:} In case where some IoT devices malfunction and stop reporting data, the cloud should still be able to obtain acceptable levels of aggregated data.
	
	\item {Efficiency:} The computational cost at each system element should be as little as possible.
\end{itemize}

\section{Preliminaries}
In order to better explain the proposed schemes, we first introduce the general process of truth discovery and cryptographic tools, in the following parts.

\subsection{Truth Discovery}
Truth discovery in large scale sensory data has been widely studied in the past. Although the algorithmic details of different solutions are a bit different from each other, the fundamental principle of assigning device weights and estimating ground truth is same. At the initialization point of truth discovery algorithm, random ground truths are assigned, which are iteratively updated until convergence is achieved. Algorithm 1 shows the general truth discovery process.

\textbf{Weight Update:} In this step, the ground truth of each object is assumed to be fixed. Typically, a device is assigned higher weight if it provides data, which is closer to the ground truth, and vice versa. Inspired by the works of CRH \cite{LiLGZFH14} (as it gives good practical performance), we calculate weight as follows:

\begin{equation}
\begin{aligned}
w_k = log(\frac{\sum^K_{k=1}\sum^M_{m=1}d(x^k_{m},x^*_m)}{\sum^M_{m=1}d(x^k_m,x^*_m)})
\label{e1}
\end{aligned}
\end{equation}
where $d(\cdot)$ is a distance function utilized to measure the difference between the ground truth and observation by devices. Moreover, $d(\cdot)$ is dependent on application use case. The two most common type of data (i.e. continuous and categorical) are considered in this work.

In applications, such as environmental monitoring, sensory data (e.g., temperature, humidity, etc.) is continuous in nature. Hence the following distance function is adopted:

\begin{equation}
\begin{aligned}
d(x^k_m,x^*_m)=\frac{(x^k_m-x^*_m)^2}{std_m}
\label{e2}
\end{aligned}
\end{equation}
where $std_m$ is used to represent the standard deviation of all the users' observations for object $m$.

Other use cases like public opinion polls have collected data that is categorical in nature, that is based on the selection of choices. In these applications, only one is correct among the multiple candidate choices. Thus, an observation vector $x^k_m = (0,\ldots,\underset{q}{1},\ldots,0)^T$ is defined to denote that the $k$-th device selects the $q$-th candidate choice for object $m$. The following function is used to measure the distance between the observation vector and the ground truth vector:

\begin{equation}
\begin{aligned}
d(x^k_m,x^*_m)= (x^k_m-x^*_m)^T(x^k_m-x^*_m)
\label{e3}
\end{aligned}
\end{equation}

\textbf{Truth Update:} In this step, weights are assumed to be fixed. We calculate the ground truth for $m$-th object as follows:

\begin{equation}
\begin{aligned}
x^*_m \leftarrow \frac{\sum^K_{k=1}w_k\cdot x^k_m}{\sum^K_{k=1}{w_k}}
\label{e4}
\end{aligned}
\end{equation}

$x^*_m$ is considered ground truth, if data is continuous. Contrary to this, $x^*_m$ is considered a probability vector where each element represents the probability of a choice being true, if the data is categorical. In this case, the final ground truth is the choice with highest probability.

\begin{algorithm}[!htbp]
	\label{A1}
	\SetCommentSty{small}
	\LinesNumbered
	\caption{Truth Discovery Algorithm}
	\KwIn{Observations from $K$ devices: $\{x^k_m\}^{M,K}_{m,k=1}$}
	\KwOut{Ground truths for $M$ objects: $\{x^*_m\}^M_{m=1}$}
	Randomly initialize the ground truth $x^*_m$\;
	\For{$iteration=1,2,\cdots,iteration_{max}$}{
		\For{$k = 1,2,\ldots,K$}{
			Update device weight(see Eq.(\ref{e1}))\;
		}
		\For{$m=1,2,\ldots,M$}{
			Update ground truth (see, Eq.(\ref{e4}))}
	}
	\Return{$\{x^*_m\}^M_{m=1}$;}
\end{algorithm}

\subsection{Cryptographic Tools}
In order to perform encryption, we make use of the following algorithms.

\subsubsection{Modified Paillier cryptosystem}
A modified Paillier cryptosystem to encrypt devices' sensitive information \cite{LiuDCW16} is used to realize privacy-preserving truth discovery. This modified Paillier cryptosystem consists of the following four components:
\begin{itemize}
	\item {\em Key Generation:} Given a security parameter $\kappa$, two large safe prime numbers $p$, and $q$ are calculated as $p = 2 p' + 1$ and $q = 2 q' + 1$, where $|p| = |q| = \kappa$, $p'$ and $q'$ are also two large primes. Then, Compute $n=pq$, and $\lambda = lcm(p-1,q-1)=2p'q'$. Choose a random value $\mu \in \mathbb{Z}_{n^2}$, and a random number $x \in [1, \lambda(n^2)/2]$. Finally, the public key is set as $pk = (n, g = \mu^2 \ \text{mod} \ n^2, h = g^x)$, and the secret key is $x$.
	
	\item {\em Encryption:} Suppose there is a message $m \in \mathbb{Z}_n$ to be encrypted. Select a random value $r \in \mathbb{Z}_{n^2}$, and calculate the ciphertexts $(c_1,c_2)$ as $c_1 = g^r \ \text{mod} \ n^2$ and $c_2 = h^r(1+n \cdot m) \ \text{mod} \ n^2$.
	
	\item {\em Decryption:} Given $(c_1,c_2)$, the message $m$ can be decrypted by computing $m = \frac{c_2/(c_1)^x - 1 \ \text{mod} n^2}{n}$.
	
	\item {\em Proxy Re-encryption:} Split the secret key $x$ into two random shares $x_1, x_2$, such that $x=x_1+x_2$. Then, the ciphertexts $(c_1,c_2)$ can be partially decrypted as $(\widetilde{c_1}, \widetilde{c_2})$ by using $x_1$, where $\widetilde{c_1} = c_1$, and $\widetilde{c_2} = c_2 / (c_1)^x$ mod $n^2$. Lastly, $(\widetilde{c_1},\widetilde{c_2})$ can be decrypted using $x_2$ to recover $m$.
\end{itemize}

\subsubsection{One-way hash chain}
\begin{figure}[htb]
	\centering
	\includegraphics[width=0.8\textwidth]{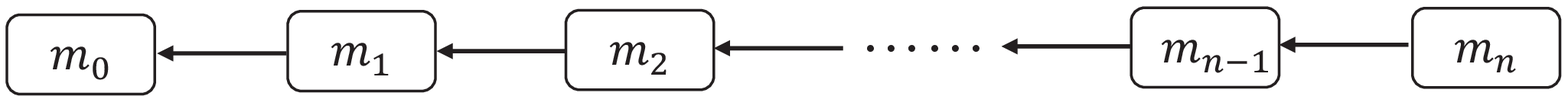}
	\caption{One-way hash chain structure.}
	\label{fig:hash}
\end{figure}

As a common cryptographic tool, various applications \cite{PerrigCTS00} have used one-way hash chain. In this work, we use this technique to authenticate the IoT devices. Suppose there is a secure hash function: $h: \{0,1\}^* \rightarrow h: \{0,1\}^l$, a one-way hash chain can be defined as a set of values $(m_0,m_1,\cdots, m_{n})$, where $m_n \in \{0,1\}^l$ is randomly chosen, and $m_i = h(m_{i+1})$ for $i=0$ to $n-1$. Note that, it is easy to compute $m_x$, where $x < y$, but becomes computationally infeasible for $m_z$, if $y < z$. Fig.~\ref{fig:hash} depicts the structure of one-way hash chain.

\subsubsection{Properties under modulo $n^2$}
In modified Paillier cryptosystem, for any message $m_i \in \mathbb{Z}_n, i = 1,2,\cdots, n$, the following equation holds
\begin{equation}
\prod^n_{i=1} (1+n \cdot m_i) \equiv (1+ n \cdot \sum^n_{i=1}m_i) \  \text{mod} \ n^2.
\label{e5}
\end{equation}

This property can be easily proven by using mathematical induction, which can be found in \cite{LuHLG17}.

\section{Proposed LPTD Schemes}
In this section, we give the details of the proposed two LPTD schemes in CIoT, which mainly include the following parts: system initialization, design overview, LPTD-I scheme, and LPTD-II scheme.

\subsection{System Initialization}
TA is considered to be fully trusted, and it bootstraps the whole system. Given a security parameter $\kappa$, TA selects two large safe prime numbers $p,q$, where $|p|=|q|=\kappa$. Following this, it then generates the public key $pk$ \& private key $sk$ of the modified Paillier cryptosystem as $pk=(n, g, h), sk = x$, where $n = p \cdot q$, and $h= g^x \mod n^2$. Then, TA randomly splits $sk$ into two shares $x_1$ and $x_2$, such that $x = x_1 + x_2$. Suppose there are $K$ IoT devices in the network, TA generates $K+2$  vectors $[s_0,s_1,\cdots,s_{K+1}]$, each contains $w$ random numbers, such that,

\begin{equation}
\sum^{K+1}_{k=0} s_{kj} \equiv 0 \mod n^2
\end{equation} 
where $j \in [1,w]$.

TA selects a secure cryptographic hash function $h$, where $\{0,1\}^* \rightarrowtail \{0,1\}^l$. Since the truth and weight are iteratively updated, we divide the number of iterations into $w$ times, and at every iteration, each device will report its observation or weighted data.  TA generates $K$ one-way hash chains $\mathcal{HC}_1, \mathcal{HC}_2, \cdots, \mathcal{HC}_K$, where $ \mathcal{HC}_k = (h_{k0}, h_{k1}, \cdots, h_{kw})$, $h_{k0} \in \{0,1\}^*$, and $h_{kj} = h(h_{k(j+1)}|| j)$, $1 \leq k \leq K$, $0 \leq j \leq w-1$.

Once these values are configured, TA assigns the keys to devices, fog node, and cloud elements, as given below:
\begin{itemize}
	\item For the device $k$, TA computes $S_k = \{(g^{s_{k1}}, h^{s_{k1}}), (g^{s_{k2}},h^{s_{k2}}), \cdots, (g^{s_{kw}},h^{s_{kw}})\}$ and assigns $S_k$, the hash chain $\mathcal{HC}_k=\{h_{k0},h_{k1},\cdots, h_{kw}\}$, and the public key $pk$.
	
	\item For the fog, TA assigns a share of the private key $x_1$, the hash chain heads of $K$ devices $(h_{10}, h_{20}, \cdots, h_{K0})$, the secret key vector $S_{K+1} = \{h^{s_{(K+1)1}}, h^{s_{(K+1)2}}, \cdots, h^{s_{(K+1)w}}\}$, the public key $pk$ , and the shared key $ss$ to the fog device.
	
	\item For the cloud, TA assigns the other share of private key $x_2$, the secret key vector $S_0=\{h^{s_{01}},h^{s_{02}}, \cdots, h^{s_{0w}}\}$, together with the public key $pk$, and the same shared key $ss$.
\end{itemize}

\subsection{LPTD Scheme: General Overview}
Once the devices obtain the observed values, LPTD will carry out the following two phases:

\begin{itemize}
	\item \textbf{Phase 1: Secure weight update.} First, every IoT device encrypts the observed value by using the cryptographic tool. Then, these ciphertexts are submitted to the fog node for aggregation and the aggregated value is further submitted to the cloud to calculate the standard deviation of the observed values, which will be then sent to every device. After that, every device computes the distances between the observed values and the ground truths. Finally, the fog and the cloud cooperatively and iteratively update the weights.
	
	\item \textbf{Phase 2: Secure truth update.} When each device receives the aggregated differences from the fog device, they first calculate the weight, the weighted observed values, and then send them to the fog device in ciphertexts. Lastly, the fog and the cloud will calculate the ground truth $x^*_m$.
\end{itemize}

During the procedure of LPTD, all operations are executed in ciphertexts. Hence, an entity only knows its own information, and the devices' sensitive information (i.e., observed value and weights) is not leaked to other entities. 

\subsection{LPTD-I Mechanism}
In this subsection, we first describe the details of LPTD-I, which is able to protect the devices' privacy and resist external false data injection attacks.

It is important to note that the sensory data from IoT devices may not be integers, but the cryptosystem used in this scheme is defined for integer values. Thus, to deal with this problem, a parameter $T$, of magnitude 10, is utilized to round off the observed values. As an example, device $k$ gets the observed value $x^k_m$ for the object $m$. We can use $T$ to multiply $x^k_m$ as $\lfloor x^k_m \cdot T \rfloor$, and the final result can be recovered by dividing $T$.  For easy understanding in this work, all observed values and intermediate results are assumed to be preprocessed as above.

\subsubsection{Secure weight update}
\textbf{Step W1.} The cloud delivers the estimated ground truth $x^*_m$ for object $m$ to all devices. If it is the first iteration, the estimated ground truth is randomly initialized. Otherwise, it will be obtained from the previous iteration. 

\textbf{Step W2.} When the device $k$ obtains $x^*_m$, it first computes the difference between $x^k_m$ and $x^*_m$ according to Eq. \ref{e2}, and then aggregates the differences of $M$ objects as $Dist_k=\sum^M_{m=1} d(x^k_m,x^*_m)$. Before submitting $Dist_k$ to the fog node,  the device uses its secret key $S_{kj}$ to compute
\begin{equation} 
\begin{aligned}
C_{kj} =  (1 + n \cdot Dist_k) \cdot h^{s_{kj}} \ \text{mod} \ n^2,
\label{e7}
\end{aligned}
\end{equation}
and then uses the hash value $h_{kj}$ to compute
\begin{equation}
mac_{kj} = h(C_{kj} || h_{kj}),
\end{equation}
where $j$ denotes the iteration number. After that, the device submits $(C_{kj},h_{kj},mac_{kj})$ to the fog. The operation may not seem time efficient, but they can be efficiently executed, as $h^{s_{kj}}$ has been calculated by TA in advance. 

\textbf{Step W3.} After receiving $(C_{kj}, h_{kj},mac_{kj})$ in the $j$-th iteration, the fog node checks the validity of the IoT device, and aggregates the reports as follows:
\begin{itemize}
	\item Check hash chain node $h_{kj}$: Assume that the fog has authenticated $h_{k(j-1)}$ in the previous $(j-1)$-th iteration, it can easily verify $h_{kj}$ according to $h_{k(j-1)} \overset{?}{=} (h_{kj} || j)$. If it holds, $h_{kj}$ is accepted. Otherwise, it is rejected.
	\item Check $mac_{kj}$: If $h_{kj}$ is valid, the fog node further verifies $mac_{kj}$ by computing 
	\begin{equation}
	mac^{'}_{kj} = h(C_{kj} || h_{kj}),
	\end{equation}
	and checking if $mac^{'}_{kj} \overset{?} {=} mac_{kj}$. If it holds, $mac_{kj}$ is accepted. Otherwise, it is rejected.
	\item  Data aggregation: After receiving $(C_{1j},C_{2j},\cdots, C_{Kj})$ from all devices, the fog node utilizes its secret key $S_{(K+1)j}$ to obtain the aggregated result as
	\begin{equation}
	C_j = \prod^K_{k=1} (C_{kj}) \cdot h^{s_{(K+1)j}} \ \text{mod} \ n^2,
	\label{e10}
	\end{equation}
	
	and then use the shared secret key $ss$ to compute
	\begin{equation}
	mac_j = h(C_j || j || ss).
	\end{equation}
\end{itemize}

Following this, the fog device delivers $(C_j,mac_j)$ to the cloud. 

\textbf{Step W4.} Upon receiving $(C_j,mac_j)$ in the $j$-th iteration, the cloud first checks data validity according to $mac_j = h(C_j || j || ss)$. If it holds, the cloud executes the following operations to obtain the aggregated results.
\begin{itemize}
	\item The cloud utilizes its secret key $S_{0j}$ to compute
	\begin{equation} 
	\begin{split}
	C^{'}_j &= C_j \cdot h^{s_{0j}} \mod n^2 \\
	&= (\prod^K_{k=1}C_{kj} )\cdot h^{s_{0j}+s_{(K+1)j}}  \mod n^2  \\
	&= (\prod^K_{k=1}(1+n \cdot Dist_k) \cdot h^{s_{kj}})  \times h^{s_{0j}+s_{(K+1)j}} \mod n^2 \\
	&= \prod^K_{k=1} (1+n \cdot Dist_k) \cdot \prod^{K+1}_{k=0} h^{s_{kj}} \mod n^2 \\
	&= (\prod^K_{k=1}(1+n \cdot Dist_k) \cdot h^{\sum^{K+1}_{k=0}s_{kj} \rightarrow 0}  \mod n^2 \\
	&=\prod^K_{k=1}(1+n \cdot Dist_k) \mod n^2 \\
	& \overset{\text{from Eq. \ref{e5}}}{\longrightarrow} \\ 
	&= 1+n \cdot \sum^K_{k=1} Dist_k \mod n^2. \\
	\end{split}
	\label{e12}
	\end{equation}
	\item The cloud can obtain $\sum^{k}_{k=1}Dist_k$ by computing
	\begin{equation}
	sum_{d} = \sum^K_{k=1} Dist_k = \frac{C^{'}_j -1}{n}.
	\end{equation}
	The cloud then selects a random number $r_{j1} \in \mathbb{Z}_{n^2}$ to blind $sum_d$ as $\log (r_{j1} \cdot sum_d)$ before forwarding it to the fog node.
\end{itemize}
\textbf{Step W5.} After receiving $\log (r_{j1} \cdot sum_{d})$, the fog node selects a random number $r_{j2} \in \mathbb{Z}_{n^2}$, and computes 
\begin{equation}
\begin{aligned}
\log(\widetilde{sum_d}) &= \log(r_{j1} \cdot sum_d) + \log(r_{j2}) \\
&=\log(r_{j1} r_{j2} \cdot sum_d).
\end{aligned}
\end{equation}

After that, the fog delivers $\log(\widetilde{sum_d})$ to the device. The device can calculate its weight as
\begin{equation}
\begin{aligned}
w_k &= \log(\widetilde{sum_d}) - \log(Dist_k) \\
&= \log(r_{j1} r_{j2} \cdot \sum^K_{k=1} Dist_k) - \log(Dist_k) \\
&=\log(\frac{r_{j1} r_{j2} \cdot \sum^K_{k=1}Dist_k}{Dist_k}) \\
&=r_j \cdot w_k,
\end{aligned}
\end{equation}
where $r_j = r_{j1} \cdot r_{j2}$.

As shown in Eq.~\ref{e2}, the standard deviation $std_m$ is necessary to calculate the difference between the observed value and the ground truth. Thus, it should be computed first. The calculations can be shown as follows:
\begin{itemize}
	\item The IoT device $k$ encrypts the observed value $x^k_m$ according to Eq. \ref{e7}, and forwards the ciphertexts to the fog node.
	\item On reception of ciphertexts, the fog node and the cloud cooperatively calculate $sum_m = \sum^K_{k=1} x^k_m$, and $\overline{x_m}=sum_m / K$ following the above operations, and then send $\overline{x_m}$ to all devices.
	\item The device $k$ calculates $d^k_m = (x^k_m - \overline{x_m})^2$, and encrypts $d^k_m$ before uploading it to the fog node.
	\item Upon receiving all the ciphertexts, the fog and the cloud cooperatively calculate $sum_d = \sum^K_{k=1}d^k_m$, and further obtain $std_m$ as $std_m = \sqrt{sum_d / K}$. At last, $std_m$ is forwarded to all devices.
\end{itemize}

\subsubsection{Secure truth update}
Upon updating the weights, it is time to update the ground truth. The details are shown as follows.

\textbf{Step T1.} The device $k$ calculates the weighted data as $r_j \cdot x^k_m \cdot w_k$, and then encrypts the weighted data and weight as
\begin{equation}
\begin{aligned}
\left \{ \begin{array}{rcl}
W_{kj,1} &=  (1 + n \cdot (r_j \cdot x^k_m \cdot w_k)) \cdot h^{s_{kj}} \ \text{mod} \ n^2 \\
W_{kj,2} &= (1 + n \cdot (r_j \cdot w_k)) \cdot h^{s_{kj}} \ \text{mod} \ n^2 \\
\end{array} \right.
\end{aligned}
\end{equation}
Then, following the same operations in secure weight update, $k$ generates $mac_{kj}=(W_{kj,1} || W_{kj,2} || h_{kj})$, and uploads $(W_{kj,1},W_{kj,2}, h_{kj},mac_{kj})$ to the fog node. 

\textbf{Step T2.} After checking the data validity, the fog uses its secret key $S_{(K+1)j}$ and runs the aggregation operations according to Eq. \ref{e10}. It then uploads $(W_j,mac_j)$ to the cloud.

\textbf{Step T3.} The cloud uses its secret key $S_{0j}$, and computes $r_j \cdot \sum^K_{k=1} (x^k_m \cdot w_k)$ and $r_j \cdot sum^K_{k=1} w_k$ according to Eq. \ref{e12}. The cloud then updates the ground truth as
\begin{equation}
x^*_m = \frac{r_j \cdot \sum^K_{k=1}(x^k_m \cdot w_k)}{ r_j \cdot \sum^K_{k=1} w_k}.
\end{equation}

Note that, we only consider continuous data in the proposed scheme. Since the difference function between continuous and categorical data is different, the distance between the observed vector $x^k_d$ and the ground truth vector $x^*_d$ can be easily computed according to Eq. \ref{e3}, which can be seen as a special case in the proposed LPTD schemes.

After combining the above two procedures, the privacy-preserving truth discovery algorithm is shown in Algorithm~\ref{A2}.

\subsection{LPTD-II Mechanism}
In real-life CIoT applications, one IoT device $l$ may not submit its data in time due to malfunctions, low battery, network delay, etc. Thus, the aggregated result is not accurate based on the previous operations, because $\sum^{K+1}_{k=0} s_k \equiv 0 \mod n^2$ does not hold. To achieve fault-tolerance, we design another efficient and privacy-preserving truth discovery approach, call ed LPTD-II. In the following, we only show how to recover the aggregated results from the ciphertexts in the cloud. Other details are omitted, as they are similar to LPTD-I.

When submitting ciphertexts to the fog node, besides $C_{kj}, W_{k,1}, W_{k,2}$, the device $k$ needs to submit another ciphertext $G_{kj} = g^{s_{kj}}$ mod $n^2$. Note that, this ciphertext is also pre-computed by TA, and delivered to the fog node in advance to save computational cost and communication overhead.

After receiving $G_{kj}$ from all devices expect the device $j$, the fog node first aggregates them as 
\begin{equation}
G_{j} = \prod^{K}_{k=1, k \neq l} G_{kj},
\end{equation}
and then uses its share of the secret key $x_1$ to partially decrypt the aggregated ciphertexts as
\begin{equation}
C_{t,1} = \frac{C_j}{(G_j)^{x_1}} \ \text{mod} \ n^2.
\end{equation}
The cloud further computes 
\begin{equation}
C_{t,2} = \frac{C_{t,1}}{(G_j)^{x_2}} \ \text{mod} \ n^2
\end{equation}
with $x_2$, and obtains the aggregated result $M$ by calculating
\begin{equation}
M = (\frac{C_{t,2}-1}{n}) \ \text{mod} \ n^2.
\end{equation}

\begin{algorithm*}[!htb]
	\label{A2}
	\SetCommentSty{small}
	\LinesNumbered
	\caption{Privacy-Preserving Truth Discovery Algorithm}
	\KwIn{Observations from $K$ devices: $\{x^k_m\}^{M,K}_{m,k=1}$}
	\KwOut{Ground truths for $M$ objects: $\{x^*_m\}^M_{m=1}$}
	The cloud randomly initializes the ground truth $\{x^*_m\}^M_{m=1}$. \\
	Each device encrypts the observed value $x^k_m$ as $Enc(x^k_m)$ and $Enc{(x^k_m)^2}$, and sends both to the fog. \\
	After receiving all ciphertexts, the fog cooperates with the cloud to calculate the standard deviation $std_m$ for object $m$, and delivers it to all devices. \\
	\For{$iteration=1,2,\cdots,iteration_{max}$}{
		\For{$k = 1,2,\ldots,K$}{
			Each device calculates the difference between $x^k_m$ and $x^*_m$, and the sum of differences for $M$ objects $Dist_k$. Then, $Dist_k$ is encrypted as $Enc(Dist_k)$, and submitted to the fog node.\\
			After obtaining $Enc(Dist_k)$, the fog cooperates with the cloud to recover $\log(sum_d)$, and further blind $\log(sum_d)$ by choosing two random values $r_{j1}$ and $r_{j2}$. Then, $\log(\widetilde{sum_d})$ is delivered to all devices. \\
			After obtaining $\log(\widetilde{sum_d})$ , each device calculates its weight, and weighed data. Both of them will be uploaded to the fog node after encryption. \\
		}
		\For{$m=1,2,\ldots,M$}{
			When the fog receives $Enc(x^k_m \cdot w_k)$ and $Enc(w_k)$, it cooperates with the cloud to calculate the ground truth $x^*_m$, and then sends the truth to all devices. \\
		}
	}
	\Return{The ground truths $\{x^*_m\}^M_{m=1}$;}
\end{algorithm*}

\section{Security Analysis}
The security properties of proposed LPTD schemes are of prime importance. Here, we show how the proposed schemes can achieve privacy preservation and effectively defend against false data injection attacks.

{\em Defense against false data injection:} To authenticate the validity of data in each iteration, one-way hash chain technique is applied in the LPTD schemes. For each device, if the hash value $h_{k(j-1)}$ is authenticated in the $(j-1)$-th iteration, $h_{kj}$ can be authenticated according to $h_{k(j-1)} = h(h_{kj} || j)$ as it is hard to obtain $h_{kj}$ from $h_{k(j-1)}$ due to the properties of one-way hash function. In fact, only if a device reports its data in the $j$-th iteration, the fog can get a fresh $h_{kj}$. If the $h_{kj}$ is not fresh in the $j$-th iteration, it can be considered as false data by replaying $h_{kj}$. The fog can identify and filter this data. Thus, the proposed LPTD schemes can defend against the false data injection attack.

{\em Privacy preservation:} In LPTD schemes, the observed value of a device $k$ is encrypted as $C_{kj} = (1+ n \cdot \overline{m}) \cdot h^{s_{kj}}$, if we look at $Dist_{kj}$ as a message $\overline{m}$. Note that $(1+ n \cdot \overline{m}) \cdot h^{s_{kj}}$ is a valid Paillier ciphertext. An external attacker cannot get $\overline{m}$, as the Paillier encryption achieves IND-CPA (i.e., indistinguishable under the chosen plain text attack). The fog node is also curious about $\overline{m}$. However, without knowing the other share of the secret key $x_2$, it will not be able to recover the sensitive data. For the weight information, $x^k_m \cdot w_k$ and $w_k$ are encrypted as $W_{k,1}$ and $W_{k,2}$ respectively. As $W_{k,1}$ and $W_{k,2}$ are both Paillier ciphertexts, an external attacker cannot recover the weight information. Notice that, the attacker may perform the following operation to calculate the weight,
\begin{equation}
\begin{aligned}
\frac{W_{kj,1}}{W_{kj,2}} = \frac{1+n \cdot (r_j \cdot x^k_m \cdot w_k)}{1+n \cdot (r_j \cdot w_k)}.
\label{e22}
\end{aligned}
\end{equation}
However, since $x^k_m$, $w_k$, and $r_j$ are unknown, the attacker cannot calculate them from Eq. \ref{e22}. The attacker may build more equations to recover $x^k_m$ as 

\begin{equation}
\begin{aligned}
\left \{ \begin{array}{rcl}
&\left \{ \begin{array}{rcl}
W_{k1,1} = (1+n \cdot (r_1 \cdot x^k_m \cdot w_{k1})) \cdot h^{s_{k1}} \ \text{mod} \ n^2  \\
W_{k1,2} = (1+n \cdot (r_1 \cdot w_{k1})) \cdot h^{s_{k1}} \ \text{mod} \ n^2 \\
\end{array} \right.
\\
\\
&\left \{ \begin{array}{rcl}
W_{k2,1} = (1+n \cdot r_2 \cdot x^k_m \cdot w_{k2}) \cdot h^{s_{k2}} \ \text{mod} \ n^2  \\
W_{k2,2} = (1+n \cdot r_2 \cdot w_{k2}) \cdot h^{s_{k2}} \ \text{mod} \ n^2 \\
\end{array} \right.
\\
& \cdots
\\
\end{array} \right.
\label{e23}
\end{aligned}
\end{equation}
From Eq. \ref{e23}, we can see that with more equations introduced, more random numbers (i.e., $r_j$) will be introduced. Since $r_j = r_{j1} \cdot r_{j2}$, only if the fog node colludes with the cloud, the attacker can obtain $r_j$. Nevertheless, under our security model, there is no collusion between the fog and the cloud. Hence, the scheme preserves the privacy, and passes the security model.

\section{Performance Analysis}
In addition to security model evaluation, we also perform experimental evaluation for communication and computational costs of both proposed schemes.
 
\subsection{Communication Overhead}
To show the communication overhead of LPTD, we compare the proposed schemes with the PPDP \cite{MiaoJSLGQXGR15}, which encrypts the data by calculating $c = g^mr^n$ mod $n^2$, under the same setting. Here, we assume the bit length of $|n^2|$ is set as $U$. However, we omit the cost of authentication for all schemes as a fairness consideration. During the process of weight update in LPTD-I, each device needs to submit $Enc(Dist_k)$, which costs $U$ bits. In PPDP,  $k$ needs to submit $Enc(Dist_k)$ and $Enc(\log(Dist_k))$, which cost $2U$. In the procedure of truth update, PPDP and LPTD-I need to submit $M  \cdot Enc(w^k_m \cdot w_k)$ and $Enc(w_k)$, which cost $(M+1)U$, where $M$ is the number of objects. Compared with LPTD-I, LPTD-II needs to submit one more $g^{s_{kj}}$ mod $n^2$ to execute the decryption operation. However, in reality, $g^{s_{kj}}$ mod $n^2$ can be submitted to the fog in advance to receive communication overhead, as it is constant. Table \ref{tab:t2} summarizes the communication overhead of all schemes in each phase for each device.

\begin{table*}[htb]
	\small
	\centering
	\caption{Comparison of communication overhead for each CIoT device.}
	\begin{tabular} {cccc}
		\toprule
		&   Phase of weight update & Phase of truth update \\
		\midrule
		PPDP  & $2U$ & $(M+1)U$   & \\ \hline
		LPTD-I   & $U$  & $(M+1)U$  & \\  \hline
		LPTD-II & $2U$ & $(M+2)U$ & \\ 
		\bottomrule
	\end{tabular}
	\label{tab:t2}
\end{table*}

\subsection{Computational Costs}
We compare the computational costs of LPTD and PPDP schemes by implementing all schemes in Java, and run several experiments on a system with 2.5 GHz Intel Core i7 and 16GB RAM. The number of iteration is set as 10, as average result of 10 experiments are used for comparisons.

\begin{figure}[htb]
	\centering
	
	\subfigure[]
	{ \includegraphics[width=0.45\textwidth, height=0.32\textwidth]{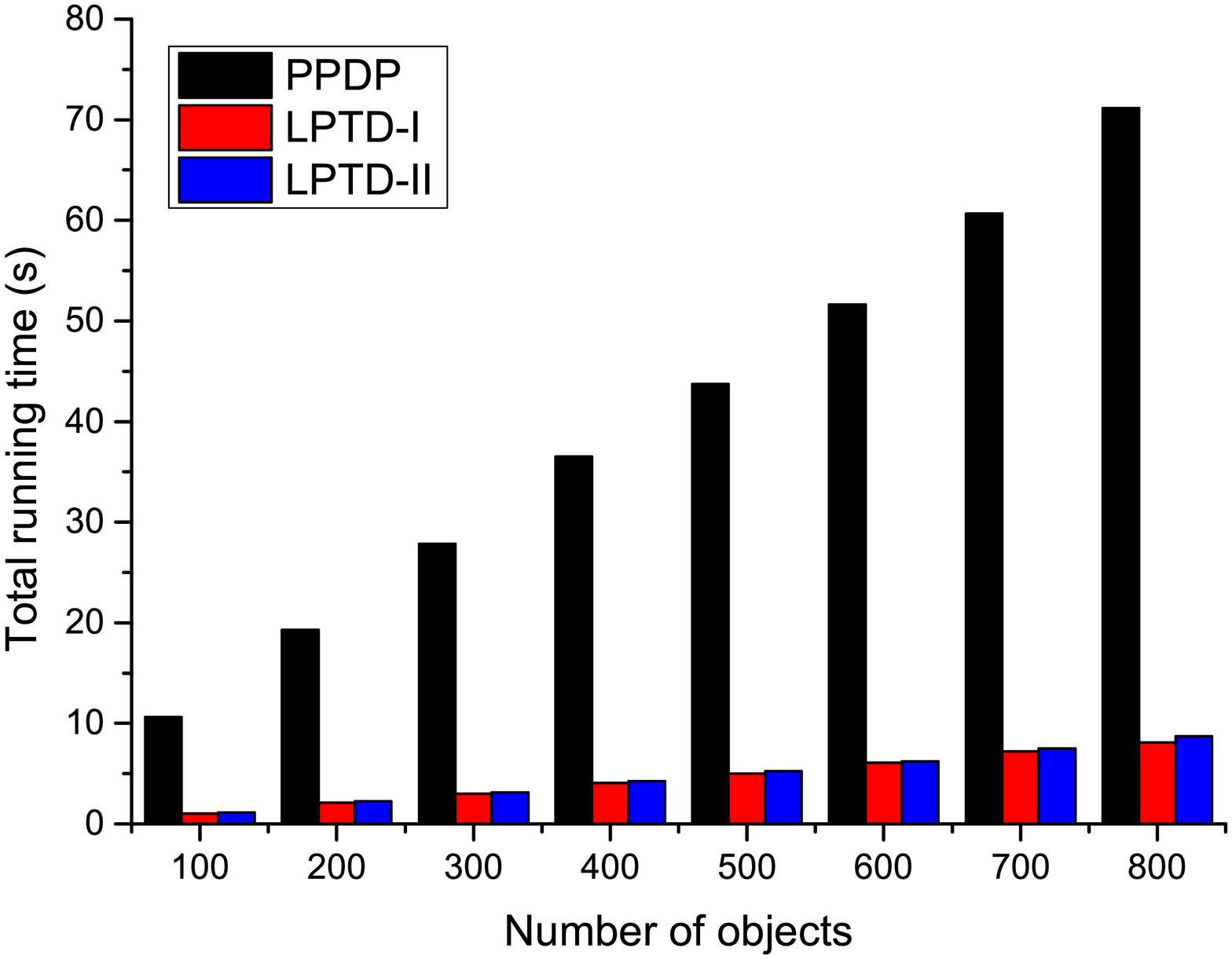}}
	~~
	\subfigure[]
	{\includegraphics[width=0.45\textwidth, height=0.32\textwidth]{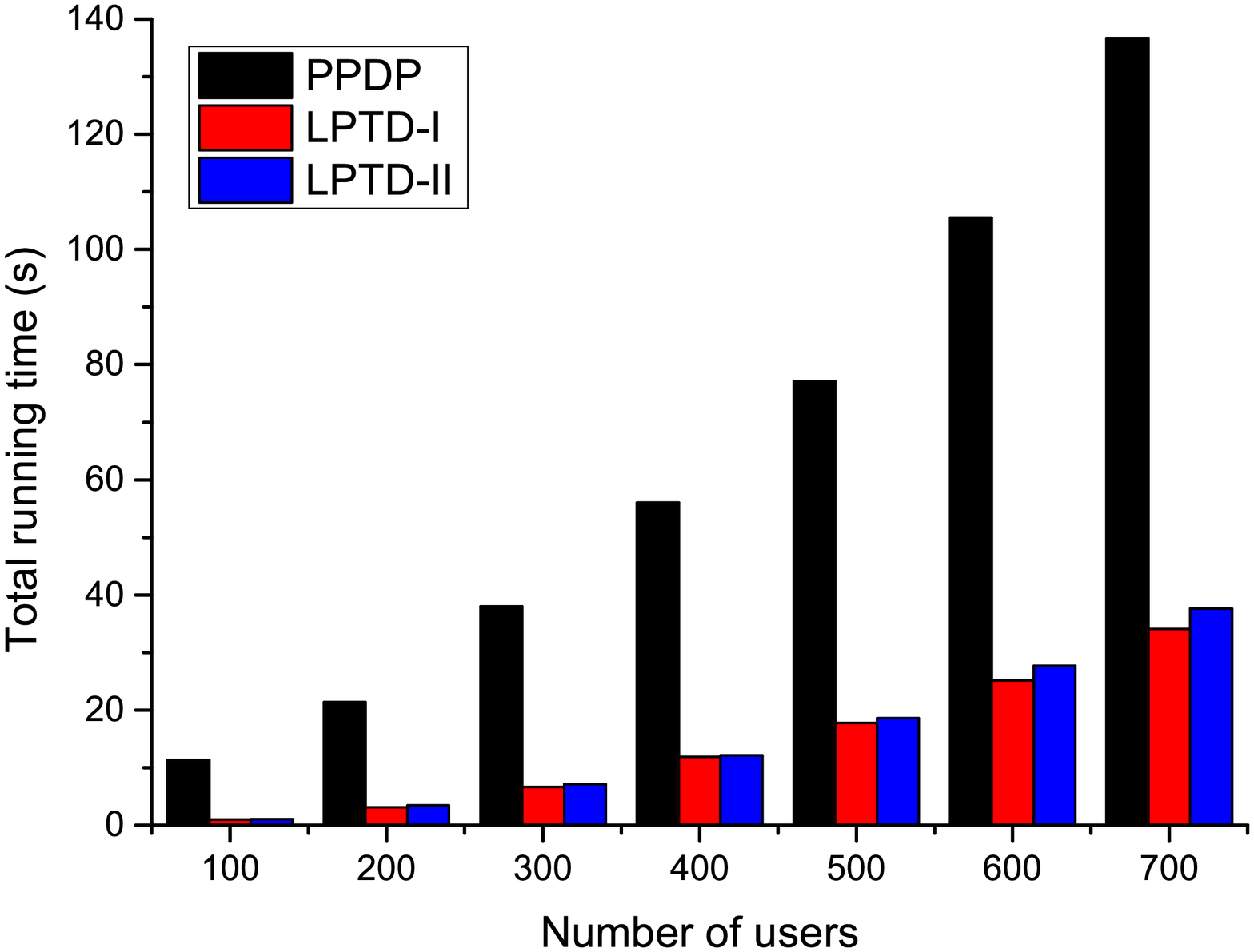}}
	
	\caption{(a) Total running time with varying number of objects. (b) Total running time with varying number of devices.}\label{fig:f2}
\end{figure}

As shown in Fig. \ref{fig:f2}(a), we compare the run time of PPDP with 100 devices and varying number of objects. It can be observed that as the number of objects increases, the run time of LPTD remains far less than that of PPDP.  For example, when the number of objects is 800, LPTD-I and LPTD-II cost 8.098s and 8.696s to finish the truth discovery respectively, while PPDP takes 71.172s. This is due to the reason that PPDP needs to perform time-consuming module exponent operations, while only multiplication operations are required in LPTD. The single module multiplication operation can be done in advance, which provides an added benefit. Note that, LPTD-I performs better than LPTD-II, since LPTD-II needs to execute 2 decryption operations to recover the aggregated results, while LPTD-I only needs to perform 2 multiplication operations. 

Similarly, from Fig. \ref{fig:f2}(b), we can also find that the total running time of LPTD is less than that of PPDP when the number of devices ranges from 100 to 700, while the number of objects is fixed at 100. When the number of devices reaches 700, LPTD-I and LPTD-II take 34.079s and 37.606s  to finish the truth discovery respectively, while PPDP needs 136.754s. This also confirms the efficiency of our scheme.

\begin{figure}[htb]
	\centering
	
	\subfigure[]
	{ \includegraphics[width=0.45\textwidth, height=0.32\textwidth]{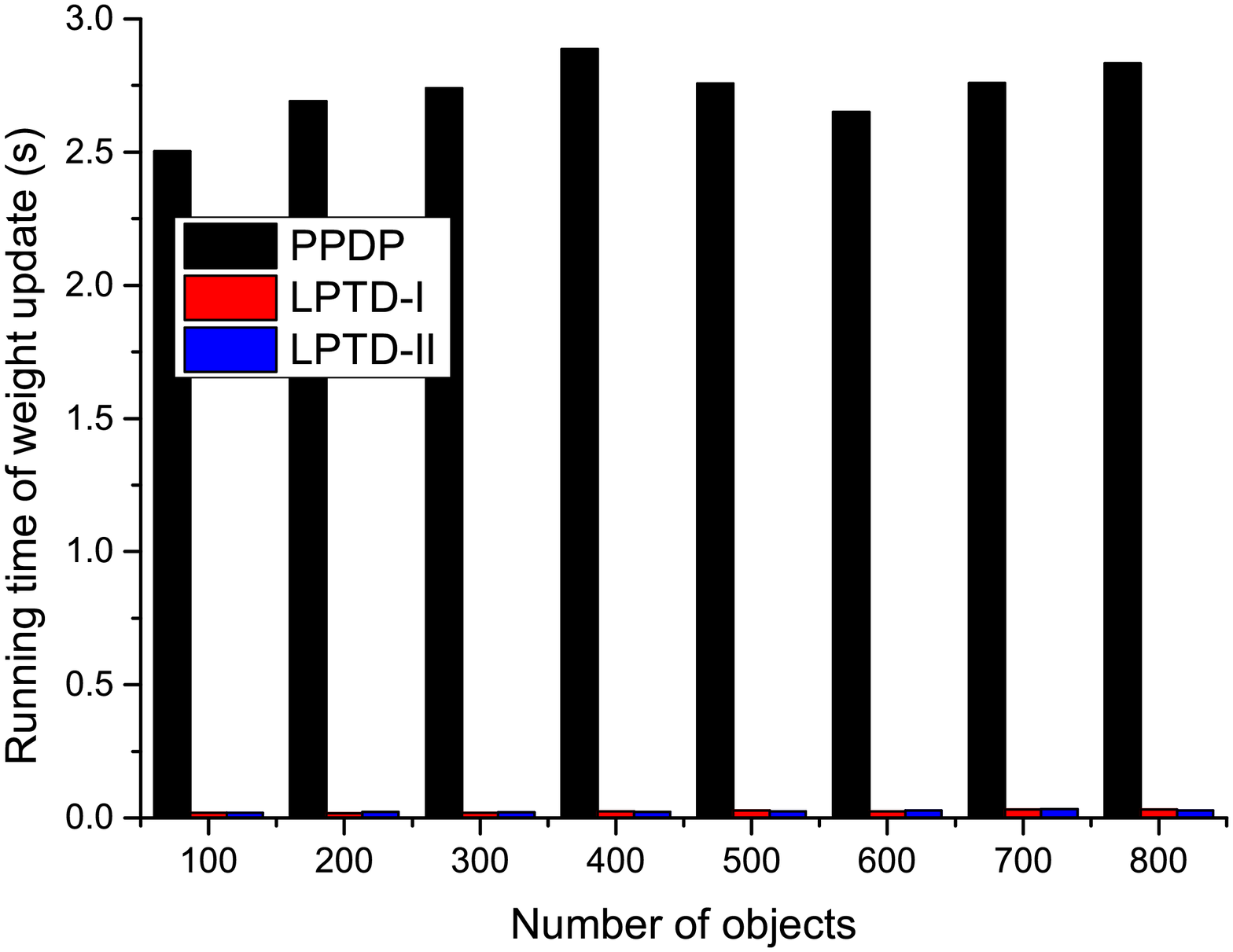}}
	~~
	\subfigure[]
	{\includegraphics[width=0.45\textwidth, height=0.32\textwidth]{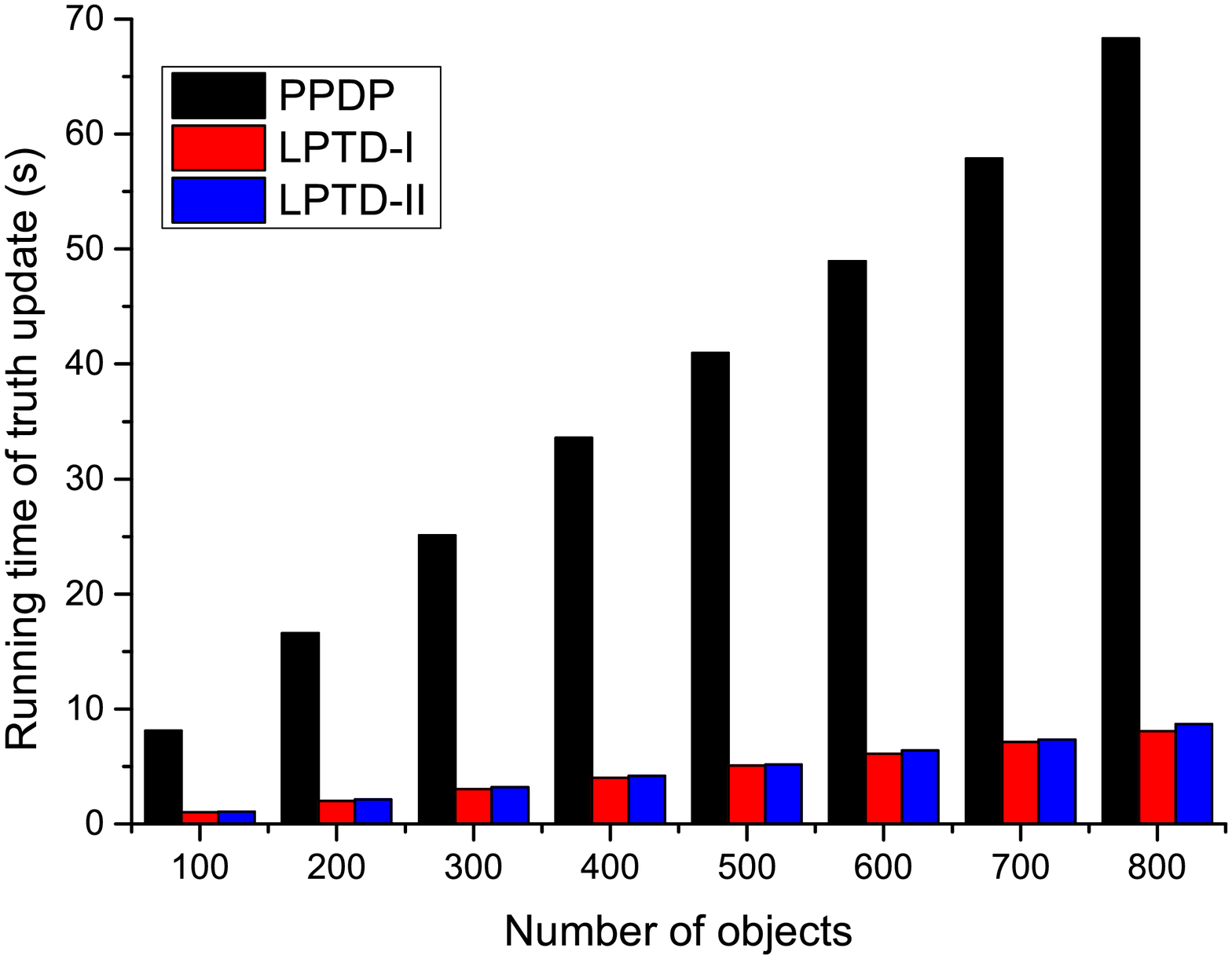}}
	
	\caption{(a) Running time of weight update with varying number of objects. (b) Running time of truth update with varying number of objects.}\label{fig:f3}
\end{figure}

Fig. \ref{fig:f3} shows the run time of weight update and truth update with varying number of objects. Here, we set the number of devices as 100. As it can be observed from Fig. \ref{fig:f3}(a), the run time of PPDP and LPTD are relatively stable. The reason is that, although more objects are introduced, each device only needs to perform 2 encryption operations in PPDP, and 1 encryption operation in LPTD (i.e., $(Enc(Dist_k), Enc(\log Dist_k))$ $vs.$ $Enc(Dist_k)$) in the weight update phase. Since PPDP needs to execute module exponent operations, it costs higher running time than LPTD-I and LPTD-II. In Fig. \ref{fig:f3}(b), the running time of all schemes grow linearly. The reason is that more truths need to be updated as the number of objects increases. It can be also found that PPDP takes higher time to finish same computations.

\begin{figure}[htb]
	\centering
	
	\subfigure[]
	{ \includegraphics[width=0.45\textwidth, height=0.32\textwidth]{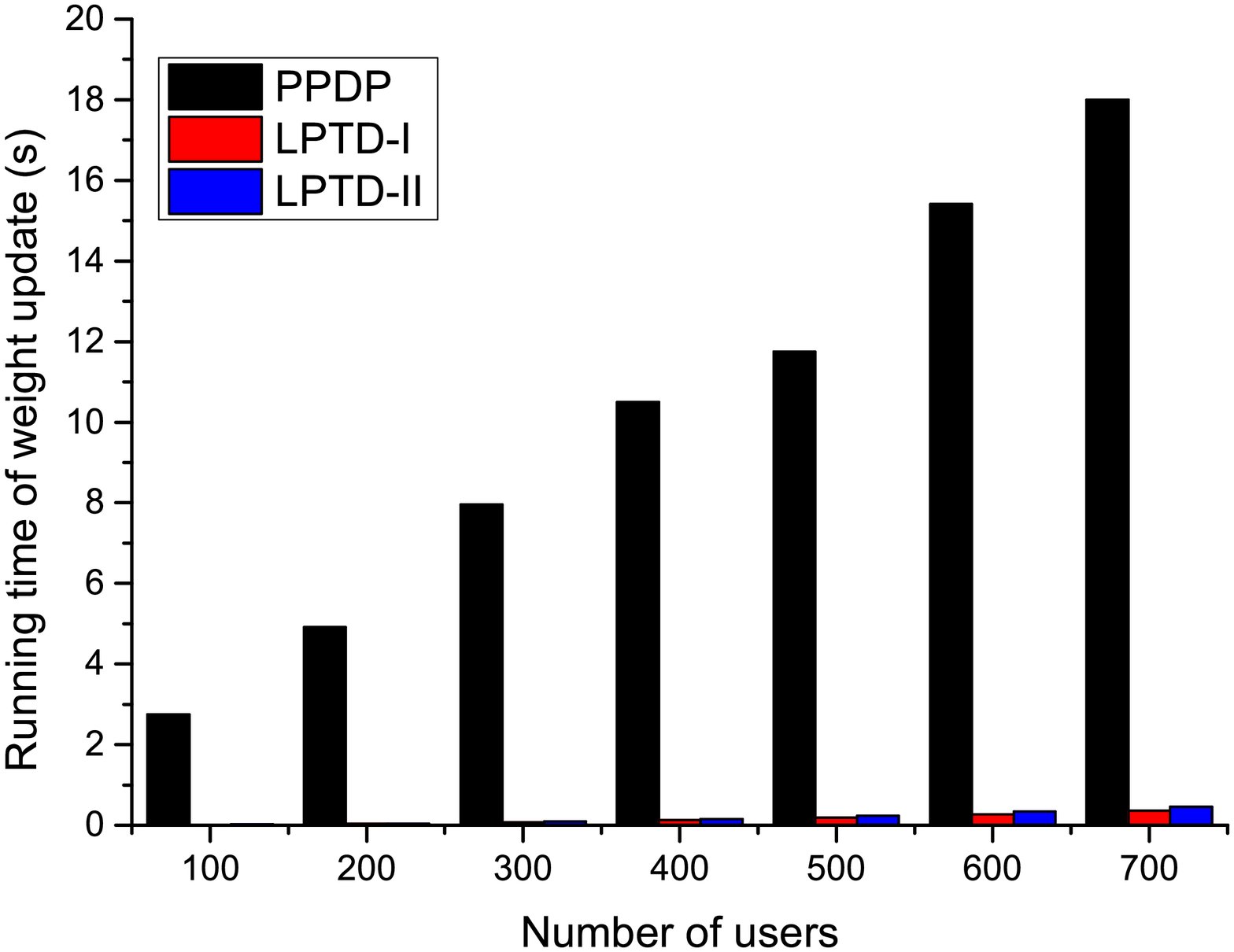}}
	~~
	\subfigure[]
	{\includegraphics[width=0.45\textwidth, height=0.32\textwidth]{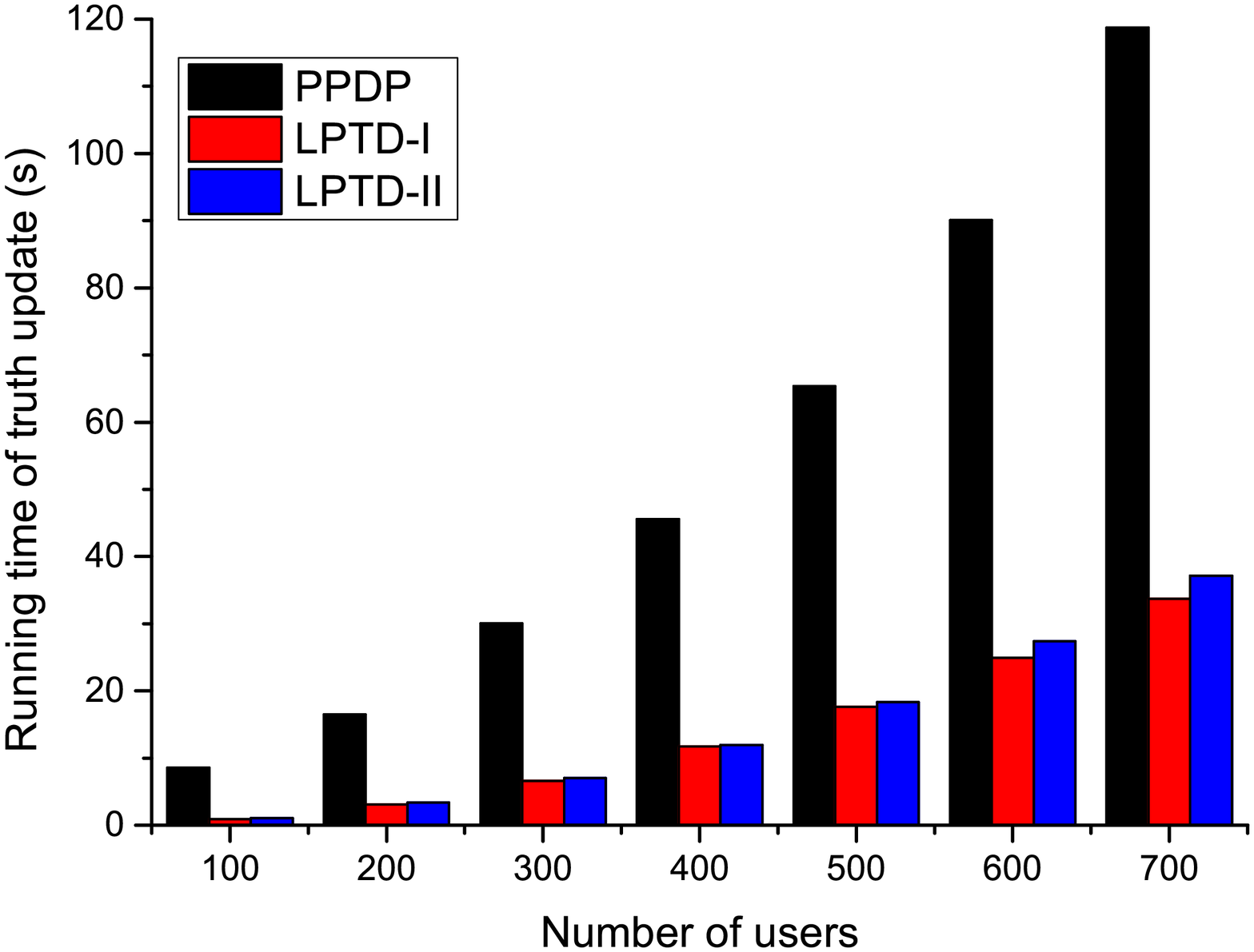}}
	
	\caption{(a) Running time of weight update with varying number of devices. (b) Running time of truth update with varying number of devices.}\label{fig:f4}
\end{figure}

Similar observations can be made in Fig.~\ref{fig:f4}. For the procedure of weight update,  since more $Dist_k$ need to be encrypted with the increasing number of devices, the run time of all schemes grows linearly.  In the procedure of truth update, as all schemes need to perform more aggregation operations to calculate $\sum^{K}_{k=1} {x^k_m \cdot w_k}$ and $\sum^{K}_{k=1} w_k$, the run time forms a linear relation with the number of devices. Based on these results, we can conclude that LPTD schemes are more efficient then existing solutions.

\section{Related work}
A number of truth discovery schemes have been studied previously \cite{LiLGZFH14,LiDLMS12,LiLGSZFH15,MaLLQGZSZJH15,MengJLGSDC15,SuLHWGLAHLGK14,WangKLA12,YinHY08,ZhangHHL12,JinSXN16}, and hence can become an attractive solution for CIoT applications. Among them, CRH \cite{LiLGZFH14}, AcuSim \cite{LiDLMS12},  TruthFinder \cite{YinHY08} are some representative schemes which can provide more reliable results by considering device reliability in the aggregation process compared to the traditional voting or averaging approaches. However, these systems fail to take into consideration important privacy issues, which may disclose some personal sensitive information \cite{DuGXC08,HeiDWH10,HeiD11}.

To protect devices' privacy, many privacy-preserving approaches have been proposed recently. For example, anonymization based schemes are presented by \cite{CramerDN01, Sweene02} to protect devices' private information. However, these cannot be used in truth discovery scenarios, since they are not designed to protect the data values. Cryptography based schemes are another option to effectively protect devices' privacy. For example, Miao et al. \cite{MiaoJSLGQXGR15} proposed a privacy-preserving truth discovery scheme by utilizing the threshold Paillier cryptosystem to protect users' privacy. However, their system is based on the assumption that there is no collusion between the cloud server and other parties. When such collusion occurs, the devices' privacy can be inferred. Moreover, cryptography schemes are not efficient, especially considering the battery and computation limitation of mobile devices. Another scheme \cite{JinSXN16} integrated the incentive with truth discovery approaches. However, the platform is trusted in their scheme which may impede its wide adoption. To improve the efficiency, Xu et al. \cite{XuLTLDY17} proposed an efficient and privacy-preserving truth discovery scheme by using an additive homomorphic data aggregation technique. Specifically, each device is assigned a random value and secret key, and the sensory data is blinded before delivering to the cloud. Finally, the authorized receivers can use the secret key and the aggregated random values to decrypt the ciphertexts. However, in real-life CIoT applications, device failure or missing data is a common issue. In such cases, this scheme does not work, since some of the random values are missing. Miao et al. \cite{MiaoSJLT17} further proposed a lightweight and privacy-preserving truth discovery scheme by using two non-colluding cloud platforms. Specifically, each device is assigned random values to perturb the sensory data, weighted data, and the weight. All these perturbed data is submitted to a cloud $S_1$, while the perturbation values are submitted to another cloud $S_2$. These two clouds can cooperatively compute the truths without disclosing the sensitive information. However, similar to \cite{XuLTLDY17}, their scheme cannot achieve fault-tolerance. Moreover, if $S_2$ eavesdrops the devices, it may decrypt the sensitive data by using the corresponding perturbation value. Finally, none of these schemes can resist external false data injection attacks.

\section{Conclusion}
This article proposes two lightweight and privacy preserving truth discovery schemes for CIoT. LPTD-I is able to use fog nodes to resist false data injections, and achieve efficient truth discovery with minimal overhead. LPTD-II is an extension to previous scheme, which in addition to attack resistance and efficient privacy preservation, provides fault tolerance. Detailed security analysis shows that the proposed LPTD schemes are secure under a comprehensive security model. Experimental evaluation shows significant reduction in computation times as compared to other schemes. 

\section*{ACKNOWLEDGMENT}
This research is supported by the National Natural Science
Foundation of China (Grant Nos. 61402037, 61272512).
\section*{References}
\bibliography{main}
\end{document}